\documentclass[lettersize,journal]{IEEEtran}
\usepackage{amsmath,amsfonts}
\usepackage{algorithmic}
\usepackage{algorithm}
\usepackage{array}
\usepackage[caption=false,font=normalsize,labelfont=sf,textfont=sf]{subfig}
\usepackage{textcomp}
\usepackage{stfloats}
\usepackage{url}
\usepackage{verbatim}
\usepackage{graphicx}
\usepackage{cite}
\usepackage{makecell}
\usepackage{color}

\begin{document}

\title{SEA Cache: A Performance-Efficient Countermeasure for Contention-based Attacks}


\author{Xiao Liu,
	Mark Zwolinski,
	Basel Halak
\thanks{The authors are with the School of Electronics and Computer Science, University of Southampton, Southampton SO17 1BJ, United Kingdom (e-mail xl8g15@southamptonalumni.ac.uk, \{mz1, bh1m10\}@soton.ac.uk).}%
}



\maketitle

\begin{abstract}
Many cache designs have been proposed to guard against contention-based side-channel attacks. One well-known type of cache is the randomized remapping cache. Many randomized remapping caches provide fixed or over protection, which leads to permanent performance degradation, or they provide flexible protection, but sacrifice  performance against strong contention-based attacks. To improve the secure cache design, we extend an existing secure cache design, CEASER-SH cache, and propose the SEA cache. The novel cache configurations in both caches are logical associativity, which allows the cache line to be placed not only in its mapped cache set but also in the subsequent cache sets. SEA cache allows each user or each process to have a different local logical associativity. Hence, only those users or processes that request extra protection against contention-based attacks are protected with high logical associativity. Other users or processes can access the cache with lower latency and higher performance. 
Compared to a CEASER-SH cache with logical associativity of 8, an SEA cache with logical associativity of 1 for normal protection users and 16 for high protection users has a Cycles Per Instruction penalty that is about 0.6\% less for users under normal protections and provides better security against contention-based attacks. Based on a 45nm technology library, and compared to a conventional cache, we estimate the power overhead is about 20\% and the area overhead is 3.4\%.
\end{abstract}

\begin{IEEEkeywords}
Cache, Contention-based attack, randomized remapping, Logical associativity, Eviction set.
\end{IEEEkeywords}

\section{Introduction}

Caches are hardware components designed to improve processor performance by reducing the access time to some recently accessed data. In a cloud computing environment, caches may be shared by different users. This resource sharing can be exploited by attackers \cite{Bhunia2019}. An attacker could retrieve other users’ data by observing cache activity. For example, in a cache-timing side-channel attack, an attacker can monitor the usage of cache lines, and hence determine if specific data was accessed. Furthermore, an attacker can deduce or retrieve sensitive data. Such a type of attack can be achieved without physical contact. Therefore, an attacker can attack cloud servers and steal useful information without being noticed by either the cloud vendor or the victims. 

Many countermeasures \cite{Qureshi2018,Qureshi2019,Werner2019,Mirage,Unterluggauer2022} have been proposed in the past decades. For example, a randomized remapping cache can protect against side-channel attacks. In this type of cache, cache lines are remapped by a dedicated function or a table. Since the attacker cannot directly predict the cache line mappings, the attack is mitigated. To enhance security, some designs provide static but overwhelming protection. They may permanently require huge power and hardware overheads \cite{Mirage}, which is not practical. Some other designs such as CEASER \cite{Qureshi2018}, and CEASER-S \cite{Qureshi2019} provide flexible protection, using a re-keying function. This type of design avoids huge permanent performance, hardware or power overheads. A privileged user can balance the performance and the security. However, stronger protections can lead to higher performance degradation, which affects all users who share the cache. Although, CEASER-S \cite{Qureshi2019} and another similar design, Scattercache \cite{Werner2019}, can still defeat a state-of-the-art contention-based attack by reducing the re-keying period \cite{P2021}, the performance degradation can become significant. Contention-based attacks have become more aggressive in recent years \cite{P2019}, and will be a greater threat within a few years.

We proposed CEASER-SH \cite{Xiao} to mitigate the performance degradation due to a high re-keying frequency. In this paper, building on the logical associativity of the CEASER-SH cache, we propose the SEA cache which can provide a better balance between cache security, performance, and hardware and power overheads. Different users can select either high protection or high performance based on their own requirements. Such a secure cache should achieve better performance for users or processes that do not require more protection than the CEASER-SH cache with a re-keying function while providing equivalent or even better protection to a protected user or processes against aggressive contention-based attacks that might be developed in the future. 


\subsection{Contributions}

In this work, we extend the logical associativity of the CEASER-SH cache to allow different users to have different logical associativity settings. Since re-keying  applies to the entire cache, all users who share the cache can be affected in terms of cache performance. Therefore, we propose the Skewed Elastic-Associativity Cache (SEA cache) to allow different logical associativity settings in different security domains. 
A high-protection domain has a higher logical associativity setting. 
Processes executed in the high-protection domain gain extra protection but some performance degradation due to the high access latency of the high logical associativity. 
Cache lines in the normal-protection domain can still be accessed with normal access latency. 
We demonstrate that the SEA cache can provide better security against contention-based attacks for specific users or processes while providing low latency and better cache performance for other users or processes.
Compared to a conventional cache, we estimate the power overhead is $0.41W$, about 20\%, and the area overhead is 3.4\%. Compared with CEASER-SH, the security is improved, in terms of reducing the success rate of contention-based attacks, by approximately 20 times. 


\section{Background}
In this section, we explain contention-based attacks and present the state-of-the-art countermeasures against such attacks.
\subsection{Contention-Based Attacks} \label{Contention-Based}
A cache is designed to reduce the latency of accessing recently used data directly from the main memory. Therefore, when a major fraction of a program resides in the cache, there is a noticeable difference in the total execution time compared with when none or just small portions of the program are stored in the cache. Therefore, the timing differences caused by cache hits and cache misses can be observed and analyzed by an attacker as a side channel. 
In a contention-based attack, the attacker prepares data that can be mapped to the same cache set as the victim's cache lines. This is known as the eviction set. 
Contentions between the attacker's and the victim's cache lines may then occur and either the attacker or the victim can evict the other's cache line. 

The most famous example of a contention-based attack is the Prime+Probe attack \cite{Osvik2006}. In the first step of the attack, the attacker primes the cache with their own eviction set. Priming can be achieved by accessing a prepared array in the attacker’s process. Then, the attacker waits for the victim's access. In the last step, the attacker re-accesses that array. If a cache set was used during the victim's process, the cache line primed in the first step would have been evicted by the victim, leading to a cache miss. After checking the all cache sets, the attacker can retrieve the cache state changes and deduce which cache sets have been accessed by the victim. \textit{Prime+Probe} provides a higher fidelity method than the other contention-based attacks, such as \textit{Evict+Time}. Also, in one round, an attacker can observe the state of more than one cache set, which improves the attack efficiency. 

\subsection{Countermeasures by Cache Remapping}
One type of countermeasure is randomized cache remapping \cite{Wang2007,Liu2016,Qureshi2018,Jaamoum2021,Zhang2021}. The main principle of this defense is obfuscating the cache line mappings, either using a table to store the cache remapping or a dedicated function to recalculate the index bits of each cache line. Since the remapping only occurs in the cache, the difference is not visible outside the cache. Hence, the distribution of the cache lines becomes much harder to predict based on the cache line address. The benefit of cache remapping is that it depends on the hardware of the cache itself, since the attacker can never manipulate the cache mapping. Although some enhancements of a randomized remapping cache need the hypervisor or kernel to be involved, the essential remapping is achieved at the hardware level. Some examples are CEASER-S \cite{Qureshi2019}, Scattercache \cite{Werner2019} and Mirage \cite{Mirage}. As secure Last Level Cache (LLC) designs, these caches use encryption ciphers, such as PRINCE \cite{Prince1}, to re-compute the addresses of cache lines. 
CEASER-S utilizes periodic re-keying, in which the cache mapping is frequently changed so that the constructed eviction set is only valid before the next re-keying. Scattercache allows all cache ways to have different mappings, namely full partitions, so that finding the eviction set that can deterministically evict the victim target cache line is more difficult. Mirage extends the V-way cache \cite{Qureshi2005} and functionally achieves a fully-associative cache, which significantly reduces the success rate of a contention-based attack. However, under such an attack with advanced profiling (see next section), the CEASER-S cache needs to re-key more frequently, which can cause huge performance degradation, and Scattercache can become vulnerable again because no re-keying is applied \cite{P2021}. Mirage does provide strong protection against contention-based attacks but requires huge hardware and power overheads.

CEASER-SH \cite{Xiao} was proposed to mitigate the performance degradation caused by re-keying to counter contention-based attacks.
It applies the idea of logical associativity to allow a cache line to be mapped to multiple consecutive cache sets. Hence, the re-keying period does not need to be as  low to provide stronger protection against an attack. Nevertheless, the high latency due to the higher logical associativity of CEASER-SH applies to all users and this can limit performance.

\subsection{Advanced Profiling} \label{AdvancedProfiling}
Constructing eviction sets, namely \textbf{Profiling}, is a crucial phase of a contention-based attack \cite{P2021}. The profiling of a conventional cache can be achieved easily \cite{Liu2015a}. However, an attack on a randomized remapping cache is more complicated for the following reasons: 1. Since the indexing is achieved by an encryption cipher, the cache mapping would appear to be random, if the cipher itself is not vulnerable.  2. If few cache ways share the same mapping, the probability of finding a cache line that has the exact mapping as the targeted cache line decreases exponentially. 3. The eviction set is only valid during a single re-keying period.

Due to these restrictions, the attacker would have to improve the profiling method to construct the eviction set using fewer cache accesses. It has been pointed out \cite{Werner2019} that, where a cache line does not share the same mapping in multiple cache ways, constructing a \textbf{Partially Congruent Eviction Set (PCE set)} on skewed randomized remapping caches is more practical. A PCE set only guarantees its member conflicts with the targeted cache line in one or more cache ways, rather than having an exact mapping in all cache ways. When applying a PCE set for a contention-based attack, the eviction of the targeted cache line is probabilistic, not deterministic. The profiling method for constructing the PCE set has been enhanced and named  \textbf{Prime+Prune+Probe (PPP)}  \cite{P2021}. 
In applying PPP, an attacker randomly selects and accesses $K$ addresses. Due to internal conflicts, some cache lines can be evicted by others. The attacker then iteratively loads those $K$ addresses into the cache or withdraws those uncached addresses, which is called Aggressive Pruning. After the victim process is triggered, the attacker re-accesses all cached cache lines and checks if one is missing. The missed cache line must have been evicted by the victim. Hence, that cache line is added to the PCE set. The attacker can finally obtain a PCE set of the expected size by repeating this process. Because the eviction is probabilistic, the more members a PCE set has, the higher the success rate.

\section{SEA cache}

\subsection{Logical Associativity}

\begin{figure}[!t]
	\centering
	\includegraphics[width=3.5in]{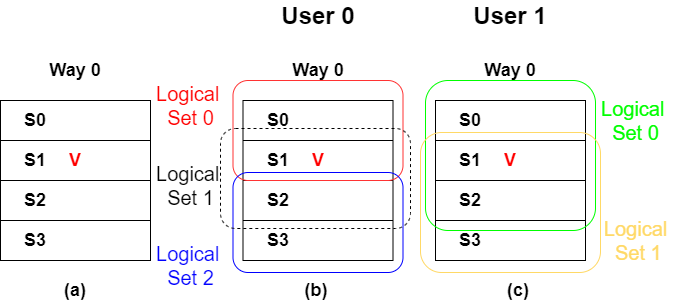}
	\caption{The alienation example of logical associativity. User 0 has a logical associativity of 2, and user 1 has a logical associativity of 3.}
	\label{Figure:Alienation}
\end{figure}

From other research \cite{P2021, Yu2019}, we note that cache associativity can affect the implementation of a contention-based cache attack. On the other hand, a high re-keying frequency in the randomized remapping cache can significantly reduce the cache performance. Therefore, we propose adaptable associativity for randomized remapping caches. By increasing the associativity and keeping the re-keying frequency relatively low, the cache can enhance its security against more aggressive contention-based attacks while the performance degradation is smaller. We name this method \textbf{Logical Associativity}. Unlike existing designs, after the encrypted index-bits are calculated, a cache with logical associativity can access both the set pointed to by the encrypted index-bits, namely the \textbf{Home Set}, and the following $H-1$ cache sets as potential placement positions. These $H$ cache lines form a logical cache set, and the value $H$ is the logical associativity number. Within this logical set, each cache line is a logical cache way. Fig. \ref{Figure:Alienation} (b) shows an example of logical associativity 2. Compared to a conventional cache in Fig. \ref{Figure:Alienation} (a), cache line $V$ can be placed in its home set $S1$ and also in $S2$. $S1$ and $S2$ form logical set 1.

We define two properties of the logical associativity of the SEA cache. The first property is \textbf{reconfigurability}. The associativity is adaptable, which  increases the attack complexity. Second, \textbf{alienation} allows different users to have different associativities. We explain these properties below. 

\subsubsection{Reconfigurability} \label{Reconfigurability}
High associativity means that, in a contention-based attack, for a given size of eviction set, the attacker can only occupy a small part of the victim's potential placement positions. However, higher associativity normally means higher cache access latency. Reconfigurability of the cache allows a privileged user, for example, a cloud vendor, to dynamically increase the associativity. Caches with self-reconfigurability have been proposed for many years. For example, in the SeReMo cache \cite{Haque2018}, each of the four cache lines can be segmented as a module, and  reconfigurability is achieved by re-setting the connections between the modules. Thus, SeReMo temporarily changes the cache's physical arrangement.

Logical associativity achieves reconfigurability in another way. We do not modify the original mapping of the randomized remapping cache but we add another module to store the logical associativity settings and send extra requests to the tag and data storage of the cache. The logical associativity can only be set by a privileged user. The privileged user can increase the logical associativity while maintaining memory coherence. Decreasing the logical associativity means flushing the entire cache or having over-ranged cache lines. 

\subsubsection{Alienation}
Alienation is the other property of logical associativity. This allows different users or processes to have different logical associativities. By providing finer-level protection, any performance degradation can be limited to the users or processes that require extra protection. Hence, this improves the overall cache performance. 

As mentioned in Section \ref{Reconfigurability}, logical associativity does not change the physical connections in the tag and data storage. Therefore, cache accesses can be made with distinct logical associativities. In Fig. \ref{Figure:Alienation}, we show the different logical associativities for different users when alienation is enabled. When the cache receives a request, the cache identifies the user by their identifier (ID). The cache can access the cache bank with the corresponding logical associativity. User 0’s logical associativity setting is 2 in Fig. \ref{Figure:Alienation} (b) and user 1’s logical associativity is 3 in Fig. \ref{Figure:Alienation} (c). Since higher logical associativity may lead to higher cache access latency, lower logical associativity can give better performance. The lowest logical associativity can be limited to a privileged user.

Based on the logical associativity, we propose the SEA cache as a defense against contention-based attacks. Alienation allows different users to have different logical associativities. Hence, they, instead of the vendor, can decide the trade-off between low logical associativity for performance or high logical associativity for security. This approach does not affect other users like changing the re-keying period would. To provide a finer balance between performance and security, we also extend the SEA cache to the process-level by using modified Security Domain Identifier (SDID) bits. This will be discussed in Section \ref{SDID}.

\subsection{SEA Implementation} \label{SubSection:SEAImplementation} 

As with other randomized remapping cache designs, we use PRINCE \cite{Prince1} as the indexing function for remapping. The tag and index bits in the cache line address are first encrypted by PRINCE. The output from PRINCE is truncated to give the encrypted index bits. For re-keying, we apply the same method as in the CEASER-S cache \cite{Qureshi2019}, which is to use two identical encryption functions but different keys. By comparing the accessed cache set and the latest remapped cache set, the cache selects the suitable encrypted index bits for the tag and data searching.

The major difference in the SEA cache is that it needs to search multiple cache sets in each cache way to implement logical associativity. Distinct from the CEASER-SH \cite{Xiao}, to minimize the latency due to accessing extra cache sets, the SEA cache accesses multiple cache banks in parallel, which is similar to PhantomCache \cite{Tan2020}. By distributing cache sets into the banks based on the last bits of the index bits, accessing consecutive cache sets can be performed as a parallel access of multiple banks. For example, the SEA cache has two banks: 0 and 1. They can only store the cache sets whose index bits end in 0 and 1, respectively. After each cache way receives the encrypted index bits, the SEA cache handles the cache access according to three different cases by comparing the number of banks ($Num_{Bank}$) and the logical associativity ($H$) of the current access:
\newline 1: \textit{$H$} is 1. \newline 2: \textit{$H$} is larger than 1 but less or equal to $Num_{Bank}$. \newline 3: \textit{$H$} is larger than $Num_{Bank}$.

In the first case, each cache way can directly send its encrypted index bits to the banks, and so behaves the same as a CEASER-S cache \cite{Qureshi2019}. In the second case, no cache sets share the same bank. Hence, all cache sets can be accessed within one round. However, except for the Home Set, all other cache sets need to be calculated by adding a different offset to the Home Set. This will cause one clock cycle delay. The accesses to the banks are only issued when all cache sets with offsets are computed. For example, if $H$ is 4, $Num_{Bank}$ is 8 and the Home set is 4, the SEA cache accesses cache sets 4, 5, 6, and 7 in one round. Thus, if a CEASER-S cache requires 43 clock cycles for a cache access, then an SEA cache requires 44 clock cycles. In the third case, since the cache sets that need to be accessed are greater than the number of banks, some cache sets share the same cache bank, thus the bank accesses are handled in different rounds. For example, if $H$ is 12 and there are 8 cache banks, the SEA cache accesses 8 cache sets in the 8 banks in the first round of access, and then accesses the remaining 4 cache sets in the second round of access. Similar to the second case, the first round requires one extra clock cycle delay to compute the offsets. In the later rounds, each round requires further cycles for the tag access. If the tag access requires 1 clock cycle, the SEA cache needs 1 clock cycle for one round of cache access. Hence, in the above example where $H$ is 12, 45 clock cycles are required.

Since the SEA cache requests multiple accesses to banks, unlike the conventional cache and CEASER-S \cite{Qureshi2019}, it needs to check multiple hit flags to determine if there is a cache hit. The SEA cache receives all the hit results of banks from each round in each cache way. Because there is either only one hit or no hits from all rounds, the SEA cache flags a cache hit when all results from each round are received. Based on the round of the cache hit and cache way, the SEA cache can track back to the actual cache set where the accessed cache line is stored. The cache set will also be sent out as a cache access result. When there is a cache miss, the SEA cache randomly selects a Home Set from one of the cache ways and adds a random offset as a replacement victim cache set. This random offset is between 0 and $H$.

\subsection{SDID and Page Sharing} \label{SDID}

\begin{figure}[!t]
	\centering
	\includegraphics[width=3in]{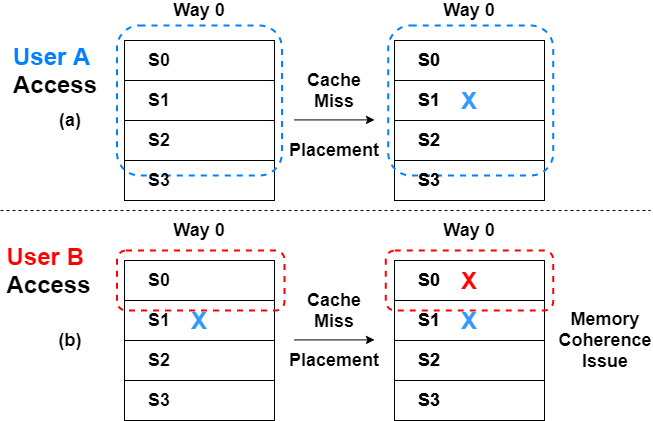}
	\caption{An example of sharing data problem while two users have different logical associativities, which finally leads to a memory coherence problem.}
	\label{Figure:LAshareError}
\end{figure}

In the SEA cache, each user can have their own logical associativity. However, having different logical associativities for each user could lead to an issue with page sharing. A user with a lower logical associativity may not be able to access the shared cache line that was accessed by another user with higher logical associativity. This can cause a fatal error in the memory system due to data coherence. For example, in Fig. \ref{Figure:LAshareError}, two users (A and B) share the same machine. 
The logical associativities for users A and B are 3 and 1, respectively. In Fig. \ref{Figure:LAshareError}(a), user A accesses the cache line X first. The home set of  cache line X is set 0. However, since the logical associativity of user A is 3, cache line X might be placed at $S0$, $S1$ or $S2$. In this example, X is placed in $S1$. Fig. \ref{Figure:LAshareError}(b) shows user B accessing this cache line X after user A's access. Due to the logical associativity of B, the cache will only search $S0$. This results in a cache miss. Hence, the cache will request another copy of cache line X with the same virtual address. This should never happen in a cache since the cache cannot distinguish between two cache lines X with the same address. If one of the cache lines has been modified, this can cause a data coherence problem. Disabling page sharing between different users is a simple solution to overcome such an issue. However, the memory system may not be used efficiently because dual or multiple copies of cache lines may be required within the same library, even for read-only access. 

To allow cache lines in a page to be shared between users, the SEA cache must guarantee that all cache lines within that page are placed with the lowest logical associativity of all users who share this page. 
Adding an identifier to each page can enable this. A similar identifier, called the security domain identifier (SDID), has been applied in some secure cache designs \cite{Werner2019, Mirage}. In the SEA cache, we use the SDID, not for different mappings, but for identifying the logical associativity of each security domain. Based on the security requirements of the user, the process can be allocated to the corresponding security domain. By default, all users’ pages are assigned SDID bits with the lowest logical associativity that the vendor sets. Hence, those processes that share the same SDID can share the same logical associativity and pages. When a user asks the hypervisor for extra protection, the hypervisor sets the SDID bits to give a high logical associativity setting in this process’s pages. Now, the SEA cache only needs to compare the SDID in the page to determine the logical associativity when accessing a cache line. 

In a contention-based attack, the attacker may not want their attacking process to be assigned the same SDID as the victim process. This is because the targeted process must be given an SDID with a high logical associativity. By having the same logical associativity, the attacking process does not gain any benefits but suffers from extra protection due to the overlapping of logical associativity. We will explain this in Section \ref{Security}. From the attacker’s perspective, they must avoid high logical associativity. 

\begin{figure}[!t]
	\centering
	\includegraphics[width=3in]{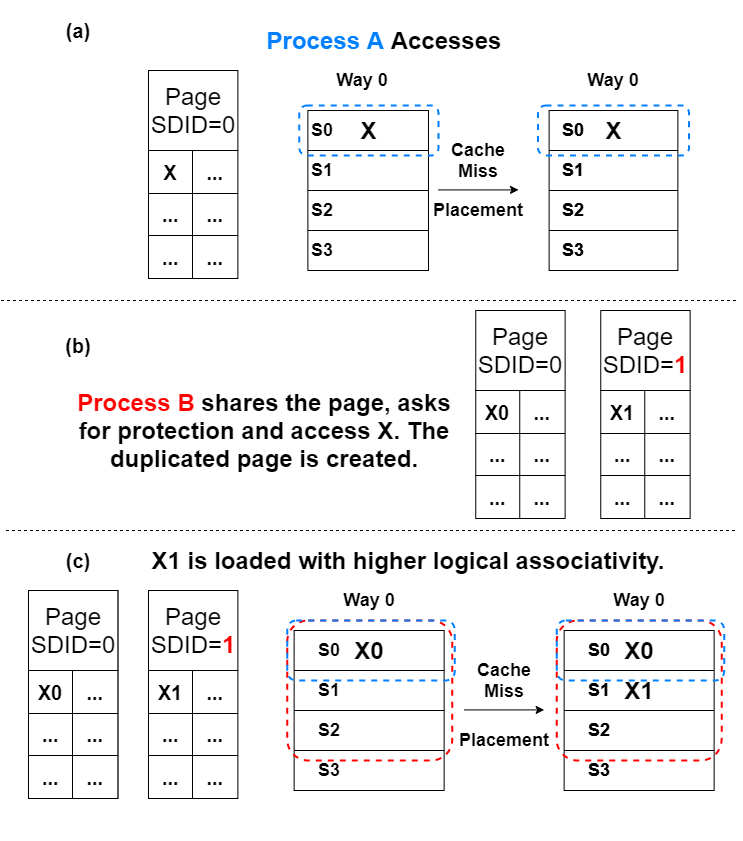}
	\caption{An example of SDID and logical associativity changes while the data is shared between A and B. Where A is in the $H=1$ domain (circled in blue) and B is in the  $H=3$ domain (circled in red).}
	\label{Figure:SEASDID}
\end{figure}

Two domains are enough for the practical implementation of the SEA cache with the SDID. One domain is set to have the lowest logical associativity, allowing all processes within this domain to achieve the lowest latency. We name it the normal protection domain. The other, high protection domain has a high logical associativity setting and is used for security-sensitive processes. Therefore, a one-bit SDID is sufficient. An example of this is shown in Fig. \ref{Figure:SEASDID}. Here, the logical associativities of the two security domains are 1 and 3. Before the sharing starts, process A accesses cache line X at the normal protection level, which is shown in Fig. \ref{Figure:SEASDID}(a). In Fig. \ref{Figure:SEASDID}(b), process B accesses the same cache line X at the high protection level and triggers the duplication of the shared page. The SDID of each page is assigned with the corresponding protection levels. After the new page is created, the cache line X1 is loaded into the cache with logical associativity 3, which is shown in Fig. \ref{Figure:SEASDID}(c). 
We only allow data sharing within the same protection level (or SDID) and disable data sharing across different protection levels. A duplicate page should be created if a page needs to be shared between two protection levels. This works as a copy-on-write, which happens when the shared data needs to be written by at least one user in a conventional hypervisor. Since most of the processes share the normal-protection level, only the security-sensitive processes request copies of the page. It should be noted that we do not consider reuse-based attacks here \cite{Liu2015a}. 

\section{Evaluation}

\begin{table*}[!t]
	\caption{The success rate of the Prime+Probe attack with optimal K (Opt K) values under different re-keying periods (RPK), logical associativities of the victim (VH) and attacker (AH).}
	\centering
	\begin{tabular}{|l|c|c|c|c|c|c|c|c|c|c|c|c|}
		\hline	
		VH/AH	& 1/1& 2/1 & 3/1 & 4/1 & 5/1 &	6/1 &8/1 &16/1 &24/1 &8/8 &8/16 &8/24\\	
		\hline	
		\diaghead{\theadfont Diag asasas}%
		{RKP}{Opt K} & 16 & 32 & 32 & 32 & 64 & 64 & 128&64&64& 32 & 64 & 16\\
		\hline
		9N & 1.00\%  & 0.53\%	&0.36\% &0.23\%	&0.23\% &	0.19\%& 0.11\%&0.06\% &0.05\% & 0.10\%	&0.04\% &0.04\%\\
		\hline
		10N & 1.17\% &0.58\%	&0.41\% &0.30\%	&0.27\% &	0.20\% &0.17\%&0.06\%&0.05\%&0.09\%	&0.06\% &0.05\%\\
		\hline
		15N & 1.78\%  &0.95\%	&0.63\% &0.43\%	&0.38\% &	0.33\%&0.24\%&0.11\% &0.08\% &0.17\%	&0.09\% &0.06\%\\
		\hline				
		20N & 2.40\%  &1.21\%	&0.81\% &0.65\%	&0.47\% &	0.44\% &0.32\%&0.16\%&0.08\%&0.24\%	&0.10\% &0.10\%\\
		\hline	
		22N & 2.80\%  &1.38\%	&0.82\% &0.67\%	&0.58\% &	0.48\% &0.35\%&0.18\%&0.14\%&0.26\%	&0.14\% &0.11\% \\
		\hline	
		25N & 3.14\%  &1.50\%	&0.97\% &0.84\%	&0.67\% &	0.58\% &0.41\%&0.19\%&0.13\%&0.26\%	&0.17\% &0.11\%\\
		\hline	
		29N & 3.68\%  &1.80\%	&1.26\% &0.93\%	&0.74\% &	0.59\% &0.48\%&0.23\%&0.15\%&0.34\%	&0.20\% &0.15\%	\\
		\hline	
		30N & 3.75\%  &1.89\%	&1.25\% &0.97\%	&0.77\% &	0.65\% &0.42\%& 0.22\%&0.18\%&0.37\%	&0.19\% &0.17\%\\
		\hline	
		35N & 4.45\%  &2.28\%	&1.41\% &1.10\%	&0.95\% &	0.80\% &0.56\%&0.30\%&0.20\%&0.40\%	&0.25\% &0.18\%\\
		\hline			
		40N & 5.05\%  &2.53\%	&1.73\% &1.22\%	&1.03\% &	 0.84\% &0.58\%&0.32\%&0.22\%&0.44\%	&0.29\% &0.19\%\\
		\hline	
		45N & 5.65\%  &2.91\%	&1.89\% &1.34\%	&1.13\% &	 0.96\%&0.72\%&0.35\%&0.29\%&0.47\%	&0.29\% &0.23\%\\
		\hline	
		50N & 6.27\%  &3.17\%	&2.11\% &1.59\%	&1.32\% &	1.03\% &0.82\%&0.41\%&0.28\%&0.60\%	&0.31\% &0.23\%\\
		\hline	
		75N & 9.63\%  &4.97\%	&3.22\% &2.43\%	&1.99\% &	1.66\% &1.25\%&0.61\%&0.44\%&0.89\%	&0.55\% &0.42\%\\
		\hline	
		100N & 12.26\%  &6.38\%	&4.15\% &3.28\%	&2.62\% &	2.17\% &1.62\%&0.75\% &0.58\%&1.17\%	&0.72\% &0.49\%\\
		\hline	
		200N & 23.15\%  &12.34\%	&8.46\% &6.35\%	&5.12\% &	4.29\% &3.30\%&1.63\%&1.11\%&2.30\%	&1.47\% &1.03\%\\
		\hline	
	\end{tabular}
			\label{Table:SEASecurity3}
\end{table*}

\subsection{Security Evaluation} \label{Security}
For the security evaluation, we modeled the SEA cache under the most aggressive contention-based attack that uses the Prime+Prune+Probe (PPP) profiling method. We wrote a simple functional simulator to model attacks. We modeled the security in two steps. In the first step, we repeated PPP until the given re-keying period was reached. The re-keying period was measured by the number of cache accesses per full cache re-keying. For example, a re-keying period of $9N$ indicates the full cache is remapped (or re-keyed) after the cache is accessed $9N$ times, where $N$ is the number of cache lines in the cache. For an $8MB$ cache with a $9N$ re-keying period, $N$ was calculated as $8 \times 1024 \times1024 \div 64 = 131,072$ (with a 64B cache line size), and the cache was fully remapped after $1,179,648$ cache accesses. During each round of PPP, the address of the evicted cache line was added to the eviction set as a member. All the addresses of the eviction set members were stored. In the second step, we took the eviction set constructed within the re-keying period to implement the Prime+Probe attack. Each eviction set was tested over $10^5$ rounds, and the overall success rate of the attacks was recorded. Hence, the success rate represents the success rate of the contention-based attack under the corresponding re-keying period. It is worth noting that we only tested the Prime+Probe attack, because the Prime+Probe attack achieves higher fidelity and is easier to implement than the Evict+Time attack. The root of both attacks is the same.

For the model, we set the following parameters. The cache size is set to 8MB, the associativity is 16 ways. The cache replacement policy was random replacement. Aggressive pruning started after the fifth round of pruning. This parameter was suggested elsewhere \cite{P2021}. In the attack, we assumed there is only one targeted victim cache line. An attacker can use all of the time for profiling before re-keying. The attacker could trigger the access of the targeted cache line itself. After triggering the victim cache line, the attacker can evict it from the cache, e.g. by flushing the entire cache. These assumptions allow the attacker to find more members of a PCE set within the re-keying period and achieve the highest theoretical success rate in a Prime+Probe attack. In other words, these assumptions are friendly to the attacker. Under such conditions, CEASER-S with full partition would need to reduce the re-keying period to about $9N$ (see above) to maintain 2-year security against contention-based attacks \cite{P2021}.

We then evaluated the security of the SEA cache with different logical associativities, using the same simulation model. Also, for comparison, we simulated the security of CEASER-SH with full partitions. We built 10 PCE sets under each configuration (300 PCE sets in total). Each PCE set is tested $10^5$ times in Prime+Probe attacks; hence each configuration was tested with one million Prime+Probe attacks. We set different numbers of initial candidate cache lines, as explained in Section \ref{AdvancedProfiling}. We set the logical associativity of the normal protection domain to 1. The highest success rates of the attack with different logical associativities for the high-protection domain, the  number of initial candidate cache line, the size of initial candidates for PPP profiling, and the re-keying periods are shown in Table \ref{Table:SEASecurity3}. $AH$ denotes the logical associativity of the normal protection domain, where the attacker is positioned. $VH$ is the logical associativity of the high protection domain, where the victim sits. $K$ is the number of initial candidate cache lines and the re-keying period is given by RKP.
The optimal $K$, from the attacker's point of view, is the value that provides the highest success rate of the attack under a particular $VH$. The lower the success rate is, the higher the protection against contention-based attacks.

In Table \ref{Table:SEASecurity3}, $AH=VH$ is equivalent in terms of security to a CEASER-SH cache with logical associativity of $H$ (the performance is not equivalent). From the table, we can see that an increase of either $AH$ or $VH$ can reduce the success rate of the attack. For example, when the re-keying period (RPK) is $200N$, the success rates of the attack are 23.15\%, 3.30\% and 2.30\% with $VH1AH1$, $VH8AH1$ and $VH8AH8$, respectively. Compared to $VH8AH1$, $VH8AH8$ has a lower success rate due to the increase of $AH$. Overall, logical associativity in the SEA cache enhances the security. For a comprehensive comparison, we also need to consider the performance. For all values of RKP, the attack success rates in the final column, $VH8AH24$, are approximately 20 times less than the rates in the first column, $VH1AH1$.

\subsection{Performance Evaluation} \label{Performance}
\subsubsection{Simulation Setup}\label{SubsubSection:SEAsetup}
To evaluate the performance of the SEA cache, we implemented both the CEASER-SH and SEA cache on the gem5 simulator \cite{Binkert2011}. We used the ARM O3 CPU model  running at a clock speed of 3GHz. The cache configurations included three levels: L1 cache with a size of 32kB and an associativity of 4, L2 cache with a size of 512kB and an associativity of 8, and L3 cache with a size of 8MB and an associativity of 16. Both L1 and L2 caches are set-associative and not shared between cores. In our simulations, we took into account the delay of PRINCE, which is 3 clock cycles, which aligns with Mirage \cite{Mirage}. We also considered the impact of logical associativity on cache latency. In the simulation, we assume there are 8 cache banks. If the logical associativity of the current access is more than one, the SEA cache requires one extra clock cycle to compute the extra cache sets. For example, when the logical associativity is one, the cache access latency is 43 clock cycles, and the cache access latency is increased to 44 clock cycles when the logical associativity is between 2 and 8. In addition, when the logical associativity exceeds a multiple of 8, the cache access latency is increased by one clock cycle. For example, when the logical associativity is set between 9 to 16, the cache access latency is increased to 45 clock cycles. Ideally, we should compare the CEASER-SH cache with the SEA cache. However, CEASER-SH does not support parallel-bank access mode. Therefore, we used the SEA cache with $VH = AH$, which is equivalent to a CEASER-SH cache with parallel-bank access mode. This can provide a much fairer comparison. 
For comparison purposes, we also tested a conventional set-associative LLC with a BIP replacement policy, which serves as the baseline in our evaluation \cite{Werner2019}. 

In the simulation, we set two workloads on two different cores. Each core was given a different logical associativity value. Core 0 is allocated to a user requiring extra protection against contention-based attacks, its logical associativity configuration, $VH$, is set to different values. Core 1 is allocated to a normal user who expects high performance and no extra protection, its logical associativity configuration is set to $AH=1$.  We used the PARSEC Benchmark Suite \cite{PARSEC} and the GAP Benchmark Suite \cite{GAP}. Since we need one workload on each of the cores, we chose two programs from the benchmark suites \cite{GAP} in each round of the simulations. 

\subsubsection{Simulation Results}\label{SubsubSection:SEAperformance}

\begin{figure}[!t]
	\centering
	\includegraphics[width=3.5in]{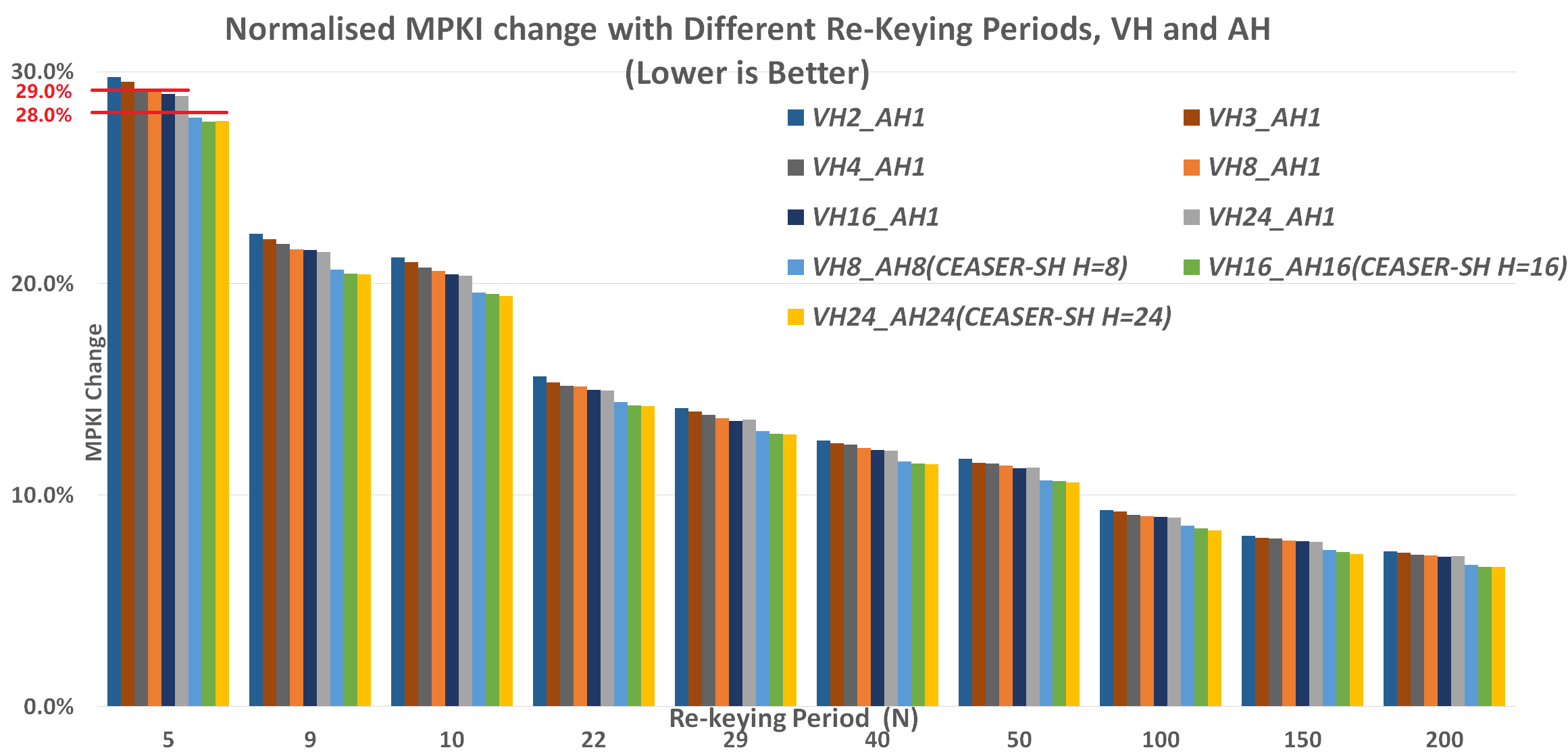}
	\caption{The MPKI of the SEA cache with different $VH$ and $AH$.}
	\label{Figure:SEAMPKI}
\end{figure}

The Misses per Kilo Instructions (MPKI) of the SEA cache simulation with different configurations is shown in Fig. \ref{Figure:SEAMPKI}. When the re-keying period is reduced, the MPKI of the SEA cache with the same $AH$ and $VH$ configurations becomes lower. Meanwhile, Fig. \ref{Figure:SEAMPKI} also shows that the SEA caches with a higher $VH$ or $AH$ lead to a slightly lower MPKI. For example, when the re-keying period is $5N$, the MPKI of the SEA cache is above 29\% when $VH2\_AH1$ and below 29\% when $VH24\_AH1$. When both the $VH$ and $AH$ are increased to 8 or more, the MPKI is further reduced to less than 28\%. This is because when only $VH$ is increased, the large logical associativity provides more potential placement positions for the cache lines from core 0 only. Furthermore, due to the cache lines from core 0 being more scattered when $VH$ is larger, the cache contentions are reduced. Hence, the overall SEA cache MPKI and the miss rate are slightly reduced when increasing the $VH$. This reduction also applies to core 1 when the $AH$ is increased. Since the miss rates follow the same trend as the MPKI, they are not shown here. 

\begin{figure}[!t]
	\centering
	\includegraphics[width=3.5in]{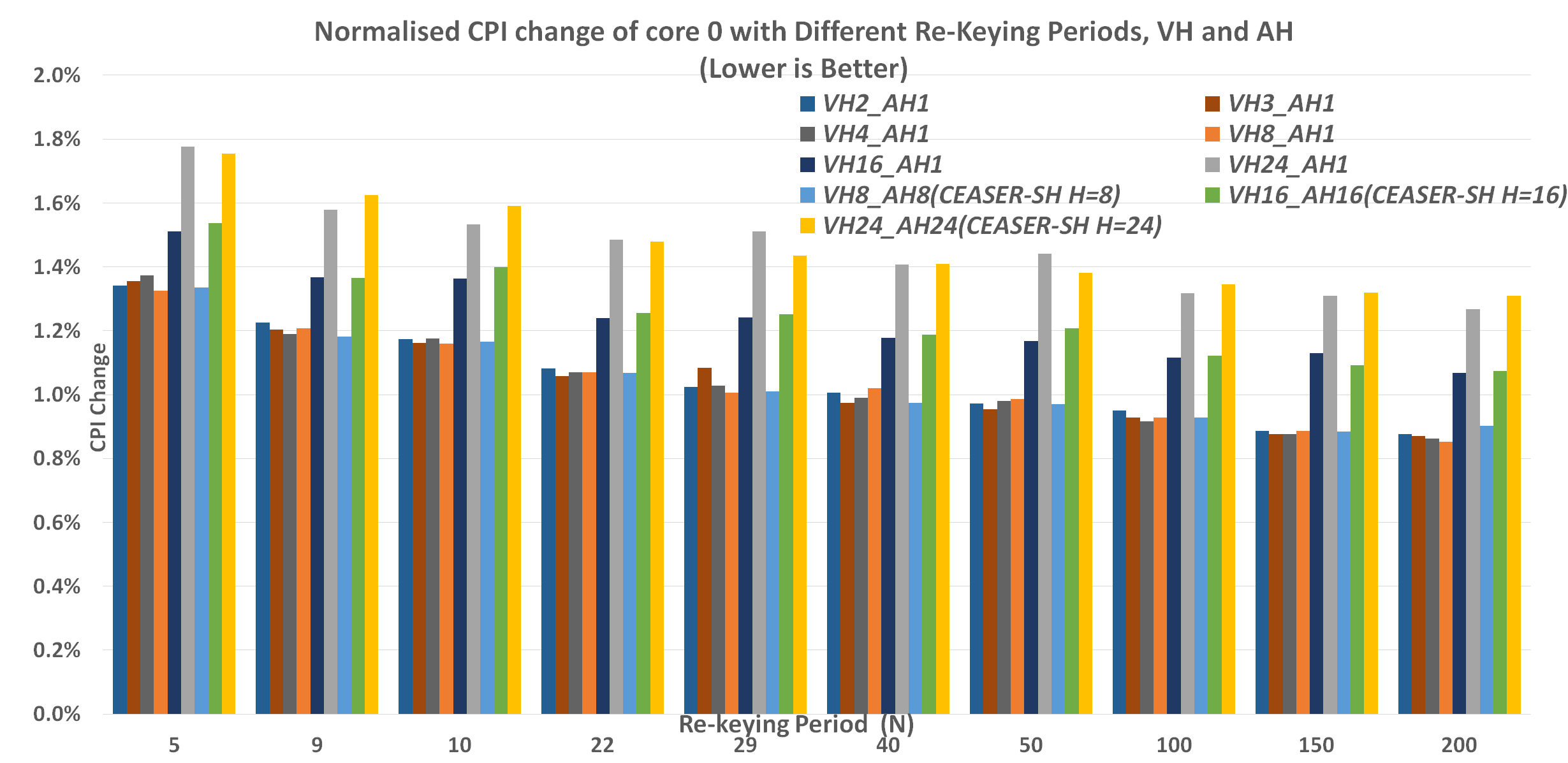}
	\caption{The CPI of core 0 for the SEA cache. Core 0 has different logical associativity, and the logical associativity of core 1 is set to 1.}
	\label{Figure:SEACPI0}
\end{figure}

\begin{figure}[!t]
	\centering
	\includegraphics[width=3.5in]{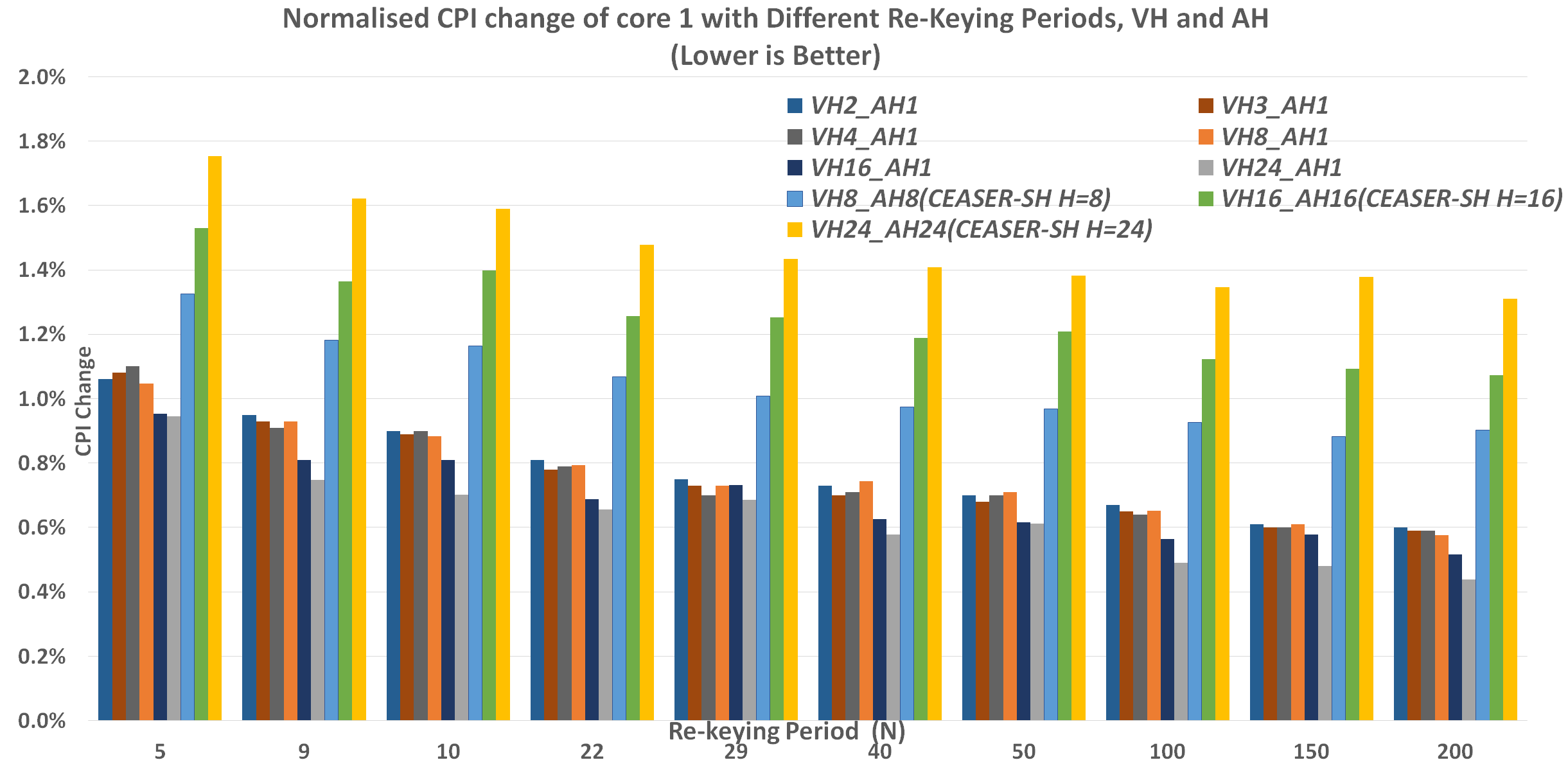}
	\caption{The CPI of core 1 for the SEA cache. Core 0 has different logical associativity, and the logical associativity of core 1 is set to 1. }
	\label{Figure:SEACPI1}
\end{figure}

The Cycles per Instruction (CPI) of both core 0 and core 1 are shown in Fig. \ref{Figure:SEACPI0} and Fig. \ref{Figure:SEACPI1}, respectively. As mentioned above, the user who requires high protection is allocated to core 0 which has high logical associativity, whereas the other user is allocated to core 1 which has low logical associativity with normal protection. From the CPI for core 0, which is shown in Fig. \ref{Figure:SEACPI0}, we can see that the SEA caches with $VH$ of 2 to 8 ($VH2\_AH1$, $VH3\_AH1$,$VH4\_AH1$, $VH8\_AH1$) have approximately the same CPI under all re-keying periods. The CPI for core 0 with $VH16\_AH1$ and $VH24\_AH1$ are always higher than for the SEA caches with $VH$ of 2 to 8. This is because the SEA caches with $VH16\_AH1$ and $VH24\_AH1$ both have a higher latency when core 0 sends the requests due to the high logical associativity. Compared with when $VH=AH$ ($VH8\_AH8$, $VH16\_AH16$, $VH24\_AH24$), we find the CPI in core 0 is highly related to the configuration of $AH$ under the same re-keying period. For example in Fig. \ref{Figure:SEACPI0}, when the $RKP=5N$, the CPI increases in core 0 are about 1.4\% when the $VH$ is 2 to 8, 1.5\%, and 1.8\% when the $VH$ is 16 and 24, respectively, irrespective of $AH$. This is because the $VH$ configuration increases the access latency by one when the $VH$ is incremented by 8 or less, where the bank number of the cache is 8.

The CPI for core 1 is shown in Fig.\ref{Figure:SEACPI1}. The CPI for core 1 with lower VH is approximately the same or even lower than the SEA cache with higher $VH$ when the $AH$ is fixed. For example, the CPI in core 1 with $VH16\_AH1$ and $VH24\_AH1$ are slightly lower than $VH$ 2 to 8. Nevertheless, such a reduction may be considered to be negligible. Compared $VH8\_AH8$, $VH16\_AH16$, $VH24\_AH24$ to $VH8\_AH1$, $VH16\_AH1$, $VH24\_AH1$, we can find all configurations with $AH=1$ achieve a lower CPI in core 1. These reductions in the CPI comes from two factors. One is the MPKI reduction, shown in Fig. \ref{Figure:SEAMPKI}, since a lower MPKI can reduce the total miss penalty. The other is that the access latency is minimum for core 1's accesses when the $AH$ is 1.

Moreover, during the evaluations, we need to consider both the cache performance and security. This is because changing the $AH$ and $VH$ can both change the cache performance or security. As an example, we consider the SEA cache with $VH16AH1$ and $AH8VH8$. From the security evaluation shown in Table \ref{Table:SEASecurity3}, when the re-keying period is $10N$, the attack success rate on the SEA cache is 0.06\%, and the SEA cache with $AH8VH8$ 
(equivalent to a CEASER-SH cache with parallel bank access mode and $H=8$) is 0.10\%, which is higher than the SEA cache.  It is worth noting that the parallel bank access mode has no effect on the security. Therefore, Table \ref{Table:SEASecurity3} is still valid.

For the performance, from Fig. \ref{Figure:SEACPI1}, when the re-keying period is $10N$, the CPI increase with the SEA cache with $VH16AH1$ is 0.8\% on core 1 (normal-protection domain). From Fig. \ref{Figure:SEACPI1}, when the re-keying period is $10N$, the CPI increase with the SEA cache with $AH8VH8$ (equivalent to CEASER-SH cache with parallel-bank access mode and $H=8$ ) is 1.2\% on core 1. As a result, for the same re-keying period, the SEA cache achieves better protection against contention-based attacks in the high-protection domain while providing better performance (lower CPI) in the normal-protection domain, even if the CEASER-SH cache also has the parallel-bank access mode.

\subsection{Hardware and Power Evaluation}
\subsubsection{Hardware Overhead} \label{SubSection:SEAHardwareOverhead}

The storage overhead of the SEA cache comes from the additional tag bits, which is the same as the ScatterCache \cite{Werner2019}. In a conventional cache, the index bits are not stored as part of the tag. The address of the cache line could be recovered directly from the tag stored and the set number of the cache. However, the set number is in the encrypted index bits. Therefore, the original index bits must be stored as part of the tag, or the encrypted index bits must be decrypted before recovering the address. Like other randomized remapping caches, we choose the former solution because the latter takes another 3 clock cycles for decryption.

We set the physical address space as 46-bits, aligned with Mirage \cite{Mirage}. We set the cache line size as 64B, the L3 cache as 8MB, and 16 ways. We compared the CEASER-SH cache, SEA cache and a conventional cache. In the conventional cache, 13 bits are used as index bits and 6 bits are offset bits. Hence, the length of the tag is 27 bits. Including 2 status bits, each cache line requires a 29-bit tag entry. There are 131072 cache lines in the cache. Therefore the size of tag storage should be $3_{,}801_{,}088$ bits which is 464kB. Both the CEASER-SH cache and the SEA cache require the full tag to be stored, which is 40 bits. Including the status bits, each cache line needs a 42-bit tag entry. The total size of tag storage is 672kB. Since we do not need to modify the data storage, the data storage of all three caches require 8192kB. The total storage size for the conventional cache is 8656kB, and  8864kB for the CEASER-SH  and SEA caches. Compared to the conventional cache, therefore, CEASER-SH and SEA caches only require 2.4 \% extra storage. Compared to Mirage \cite{Mirage}, which requires about 20\%  storage overhead, the increase is negligible. 

We have estimated the area and power overhead of the additional control logic using CACTI 6.0 \cite {CACTI} and the Nangate45nm PDK \cite{Nangate}. Our cache design is based on the SiFive L2 inclusive cache \cite{SiFive}. Compared to the area of the entire storage in the conventional cache, the area overhead is about 5.1\%. 
For  the SEA cache, we assume there are 8 cache banks. 
The bit width should be aligned with the index bits of the SEA cache. For example, for a SEA cache with 8192 cache sets, the bit width should be 13 bits. After synthesis, we found that the total area overhead including the storage is about 3.4\%, compared to the conventional cache. Even if SDID is applied, the hardware overhead is the same. This is because, as with Scattercache, the SEA cache with SDID does not require extra hardware, the SDID can be implemented via the user-defined bits in each page table entry \cite{Werner2019}. Although CEASER-SH requires no extension modules, the total area overhead is still about 3.4\%.

\subsubsection{Power Overhead Evaluation} \label{SubSection:SEAPower}

Since the logical associativity of each user is different, the dynamic energy required for each access might be different. We assume half of the cores use logical associativity 1, and others use logical associativity 16. When the logical associativity is 16, the dynamic power is 16 times the dynamic power needed for the cache access with logical associativity 1. Based on the power results from CACTI6.0 \cite{CACTI} and the synthesized circuits, we estimate the power consumption of the SEA cache. Compared to the conventional cache, SEA requires $0.41W$ extra power, which is a 20\% increase. The dynamic power due to the extra bank accesses dominates the power overhead. This overhead is approximately the same as Mirage \cite{Mirage} and much lower than the PhantomCache \cite{Tan2020}. CEASER-SH cache with logical associativity of 8, which was compared in Section \ref{SubsubSection:SEAperformance}, requires $0.39W$ power. The power increase in SEA cache is $0.02W$, which is  minor. Nevertheless, it is worth noting that the actual power overhead of SEA cache with SDID should be much less, since most processes can be executed under a normal-protection domain. 

\section{Conclusions}
Many countermeasures have been proposed to overcome contention-based attacks in the last 15 years. As countermeasures have been developed, the attacks have improved as well. To protect against more advanced attacks, most existing randomized remapping caches either require a much shorter re-keying period or use an architecture which provides sufficient protection but has permanent performance degradation. To mitigate the performance loss while protecting against contention-based attacks, we propose a new protection scheme for the randomized remapping cache, which is called Logical Associativity. It allows a cache line to be placed not only in the cache set that it maps to but also in the following consecutive cache sets. Based on the logical associativity, we propose a Skewed Elastic-Associativity (SEA) cache. In the SEA cache, different users are allowed to have their own local logical associativity settings. Hence, users who do not require extra protection do not need to suffer performance degradation caused by other users who require extra protection. For improving the SEA cache access latency, we enable parallel bank access. We also use SDID to provide finer protection to the process level, which also helps page sharing become more efficient. An SEA cache with logical associativity of 1 for normal protection users and 16 for high protection users achieves more than 2 years of protection against the strongest contention-based attack. Compared to a CEASER-SH cache with parallel-bank access and logical associativity of 8, SEA cache only requires a CPI increase of about 0.6\% for users under normal protections and increases the security, in terms of reducing the success rate of contention-based attacks by approximately 20 times. Based on a 45nm technology, compared to a conventional cache, we estimate the power overhead is $0.41W$, about 20\%,  and the area overhead is 3.4\%, which may be considered to be acceptable.

\bibliographystyle{IEEEtran}
\bibliography{SEA}

\end{document}